# MODELLING OF MEASURING SYSTEMS
# – FROM WHITE BOX MODELS TO COGNITIVE APPROACHES

*Nadine Schiering* [a*], *Sascha Eichstädt* [b], *Michael Heizmann* [c], *Wolfgang Koch* [d], *Linda-Sophie Schneider* [e], *Stephan Scheele* [f], *Klaus-Dieter Sommer* [g]

[a] ZMK & ANALYTIK GmbH, Bitterfeld-Wolfen, Germany, n.schiering@zmk-wolfen.de
[b] Physikalisch-Technische Bundesanstalt, Berlin, Germany
[c] Karlsruhe Institute of Technology KIT, Karlsruhe, Germany
[d] Fraunhofer Institute for Communication, Information Processing and Ergonomics FKIE, Wachtberg, Germany
[e] Friedrich-Alexander-University Erlangen-Nuremberg, Erlangen, Germany,
[f] Fraunhofer Institute for Integrated Circuits, Erlangen, Germany
[g] Technische Universitaet Ilmenau, Ilmenau, Germany
* Corresponding author

*Abstract* − Mathematical models of measuring systems and processes play an essential role in metrology and practical measurements. They form the basis for understanding and evaluating measurements, their results and their trustworthiness.

Classic analytical parametric modelling is based on largely complete knowledge of measurement technology and the measurement process.

But due to digital transformation towards the Internet of Things (IIoT) with an increasing number of intensively and flexibly networked measurement systems and consequently ever larger amounts of data to be processed, data-based modelling approaches have gained enormous importance.

This has led to new approaches in measurement technology and industry like *Digital Twins*, *Self-X Approaches, Soft Sensor Technology* and *Data and Information Fusion.*

In the future, data-based modelling will be increasingly dominated by intelligent, cognitive systems. Evaluating of the accuracy, trustworthiness and the functional uncertainty of the corresponding models is required.

This paper provides a concise overview of modelling in metrology from classical white box models to intelligent, cognitive data-driven solutions identifying advantages and limitations. Additionally, the approaches to merge trustworthiness and metrological uncertainty will be discussed.

**Keywords:** white box modelling; data-based modelling; measurement and information; cognitive approaches in metrology; trustworthiness and uncertainty

## 1. INTRODUCTION

Mathematical models of measuring systems and processes play an essential role in metrology and practical measurement, starting with the development of new sensors and measuring systems through to the most common task, the determination of measurement results including the evaluation of their measurement uncertainties.

The recently published "Guide to the expression of uncertainty in measurement — Part 6: Developing and using measurement models" [1] provides an overview of typical modelling approaches in metrology and their use for the evaluation of measurement uncertainty.

This paper provides a concise overview of modelling in metrology from the classic approach of *analytical parametric modelling*, and (complementary) classical *parameter-based modelling* to future-oriented *cognitive data-based modelling* approaches.

Another differentiation of models in metrology leads to so-called *white box, grey box, and black box models* [2]-[4]. In Figure 1 [2]-[3] an overview is provided of these three types of models including their advantages, disadvantages in applications of classical metrology and cognitive systems.

Figure 1. Modelling approaches in metrology and of cognitive measuring sensors [2]-[3].

Usually, a model consists of a set of mathematical equations involving at least two quantities. In general, a model can be presented by:
− *analytical models* using algebraic equations, ordinary and partial differential equations,
− *graphical models* using block diagrams, state graphs, Petri nets and
− *numerical models* using (sequences of) data with value discretisation or time discretisation.

## 2. FUTURE DEVELOPMENTS RELEVANT TO MODELLING

Digitalization, strongly and flexible networking of measuring systems, development of cognitive sensors and measuring systems, quantum metrology and sensors are just some examples of developments that are highly relevant for future modelling approaches and strategies.

In addition to process and production measurement, measurement technology and consequently modelling is faced with new fields of application including autonomous systems in road and rail transport, energy grids, earth and climate observation, remote medicine and military reconnaissance.

In this paper, we first discuss the analytical parametric modelling which is based on sound knowledge of the measuring system (section 3). We then look at classical data-based modelling which is based on processing and evaluation of measurement data; in the past mainly used in addition to analytical-parametric modelling, for example in the transition from ideal to real thermodynamic behaviour of gases (section 4). Section 5 is about input quantities, signals and their propagation. Section 6 deals with flexible, information-based networked systems and is divided into three important application areas: *Sensor Data Fusion, Digital Twins,* and *Quantum Sensors*. Section 7 covers the most important current development, i.e. modelling of cognitive measuring systems and the quantitative evolution of trustworthiness in respective measurement results. Our concluding remarks are given in section 8.

## 3. ANALYTICAL PARAMETRIC MODELLING

Classical analytical parametric modelling is an established modelling approach and based on the sound knowledge of the measurement process, of the measurand(s) and influencing quantities as well as on their relationships in cause-effect direction [2]-[4]. *Analytical parametric modelling* corresponds in its "pure" form to the *white box* models and is knowledge-based (of technology, physics, etc). Therefore, these models establish a physical relationship between the measurand and all relevant input quantities on which the output signal or indication depends.

The classical method of understanding a measuring system or process follows a path from cause to effect. In consequence, the measurand might be evaluated based on the inverse cause-effect equation (cf. Figure 2 [2][5]). This means that every measurement is an inverse problem. If it is a so-called ill-posed inverse problem, the possibilities of classical analytical modelling could be exceeded, and data-based approaches could be used to supplement it (grey box models).

The measurement equation describes the mathematical relationship between the indication, the influencing quantities and the measurand and thus leads to the measurement result. The result is a model with almost completely characterized properties, which is often referred to as a white box model. Typically, the quantities used for such modelling are tangible physical quantities such as pressure, temperature, flow rate, or angle [2].

The advantages of analytical parametric models lie primarily in the good interpretability of the model, as the quantities and function blocks that occur generally enable a concrete physical interpretation. Often, this makes it easy to check the basic model properties for plausibility. By modelling concrete physical properties, analytical parametric models often enable a good description over a large range of values of the modelled quantities - whereby the validity range must of course be considered.

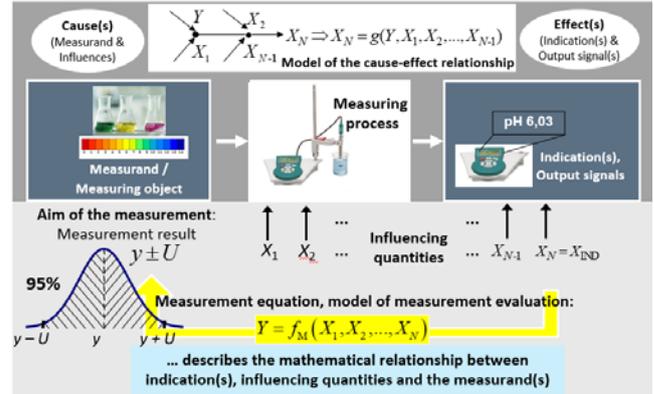

Figure 2. Cause-effect relationship and (inverse) measurement equation (modified from [2][5])

The inverse model, called measurement equation [6], or evaluation model (cf. Figure 2) [7] forms the basis for applying uncertainty evaluation methods as described in the Guide to the Expression of Uncertainty in Measurement (GUM) [7].

Figure 3 shows the process of developing an analytical parametric model using the example of determining a weighing value [8] starting from a simplified illustration and block diagram over the derivation of the cause-and-effect equation to the inversion to the model equation.

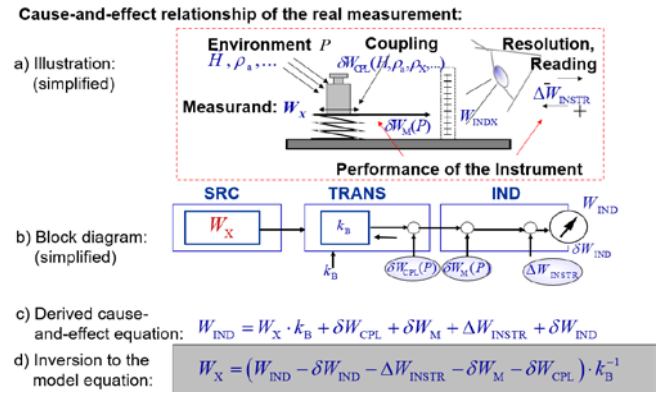

Figure 3. Example of a cause-effect model of a real measurement (determination of a weighing value [8]) and inversion to the model equation

The basis of analytical parametric modelling is the necessary physical understanding of the measurand and the measuring process, which generally requires expert knowledge or experience of the measurand and of the measuring process. This might be seen as a disadvantage of analytical parametric models. If the measurand or measuring process is not known from the outset, considerable effort may be required for the analysis and the parameterisation of the model to determine the relevant influencing quantities and cause-effect relationships or to carry out suitable experiments.

The modelling of dynamic measurements is of growing importance for industry and metrology. However, the GUM does not deal with the modelling of dynamic measurements. Dynamic measurements can be described by ordinary differential equations. The dynamic transmission systems unavoidably produce dynamic errors. These can be expressed as instantaneous dynamic error. In practice, however, this is not very useful. Therefore, dynamic mean-square errors are used. Figure 4 illustrates the modelling approach for dynamic measurements [9].

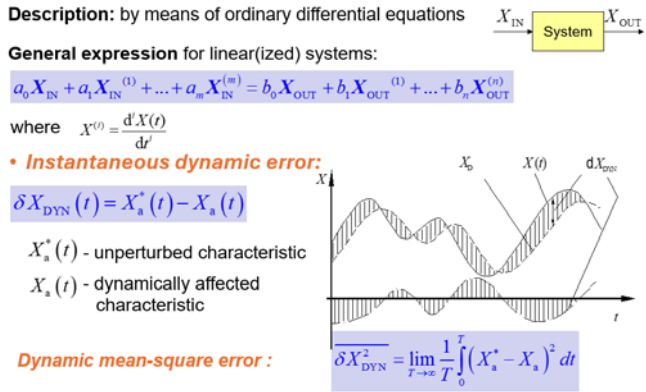

Figure 4. Illustration of the mathematical approach for dynamic measurements [9].

Classic analytical parametric modelling (white box models) still dominates metrology up until today. This applies not only to linear, time-invariant (LTI) systems, but also to time- and frequency-dependent and distributed measurement systems. In Table 1 [10] the basic equations for "classic" system approaches for analytical parametric modelling are provided.

Table 1. Equations for "classic" system approaches for analytical parametric modelling [10]

| | | General systems | Linear systems (LTI systems) |
|---|---|---|---|
| Static or time invariant systems | Algebraic equations | Equations without time dependencies $\underline{Y} = f(\underline{X})$ | Linear systems of equations $\underline{Y} = \underline{AX}$ |
| Dynamic or time depen-ded systems | Differential equations | $\underline{Y}(t) = f(\underline{Y}^{(1)}, \underline{Y}^{(2)},..., \underline{X}, \underline{X}^{(1)}, \underline{X}^{(2)},..., t)$ | Lineare differential equ. $\sum_{\mu=1}^{n} a_\mu \underline{Y}^{(\nu)} = \sum_{\mu=1}^{m} b_\mu \underline{X}^{(\mu)}$ |
| | State space models | $\underline{\dot{Z}}(t) = f(\underline{Z}(t), \underline{X}(t), t)$ $\underline{Y}(t) = g(\underline{Z}(t), \underline{X}(t), t)$ | $\underline{\dot{Z}}(t) = \underline{AZ}(t) + \underline{BX}(t)$ $\underline{Y}(t) = \underline{CZ}(t) + \underline{DX}(t)$ |
| | Description in frequency domain | | $\underline{Y}(s) = G(s)\underline{X}(s)$ $G(s) = \sum_{\mu=0}^{m} b_\mu s^\mu \left( \sum_{\nu=0}^{n} a_\nu s^\nu \right)^{-1}$ |

In some cases, the unknown complex relationships are problematic. Therefore, analytical parametric modelling is often simplified and idealised to enable an analytically explainable description or solution.

In general, if the physical, chemical and other relationships are clearly known, there is no functional uncertainty for analytical parametric models. That is, there is no additional uncertainty contribution from the model itself, but only from the uncertainties associated with the input quantities. However, it would be possible to take functional uncertainty into account if for individual parameters model approximations are necessary.

## 4. DATA-BASED MODELLING

In principle, in data-based or data-driven modelling, known structures, physical relationships and similar properties of the measuring system do not need to be analysed and mapped. In contrast, the measuring system is only described through interaction with its environment at the measurement inputs and outputs [2]-[4]. This interaction is recorded by systematically observing the input and output signals and transferred to a mathematical model. As the internal structure of the model is not known, such a model is also referred to as a black box model. In its "classical" form, data-based modelling uses well-known series and error approaches of mathematics, e.g. power series or *Taylor series* approaches. Examples are real gas law or equalisation calculation for range calibrations.

In principle, no explicit knowledge of the internal structure of the measuring system is required to create the model. This method can therefore also be used in cases where the system structure and the interaction of the components in the measuring system are not or not sufficiently known.

However, in reality, a minimum level of prior knowledge must always be available, which is why data-based models are usually grey box models.

Analytical parametric and data-based models are not necessarily in competition with each other but can rather be combined to exploit the advantages of both approaches [2][3].

Although in analytical parametric modelling, the physical structure of the system is defined and the model parameters are initially determined by analysing the system, the quality of the parameter fit can usually be improved by later data-based optimization. This approach can be used advantageously, for example, if the internal state variables of the system are known but are coupled by unknown or difficult-to-identify (e.g., strongly non-linear) interactions. In a pure white box model approach, the unknown interactions would have to be included using probabilistic arguments or as model uncertainties. A combined approach of analytical parametric and data-based modelling can reduce the overall uncertainty [2].

Figure 5 shows a data-based model using the example of the temperature dependence of pH buffer solutions using polynomial approximation.

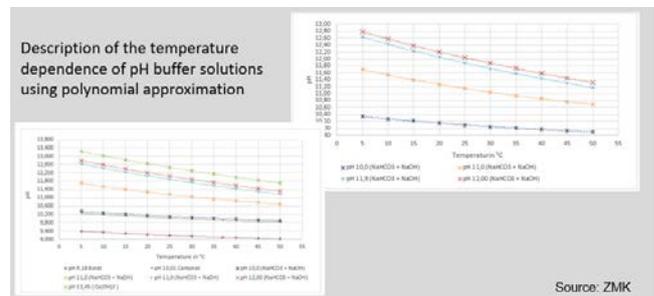

Figure 5. Example of data-based model based on the description of the temperature dependence of pH buffer solutions.

A quantitative evaluation of data-based models regarding accuracy, trustworthiness or even the functional uncertainty of the models, equivalent to uncertainty in metrology, is required. This can be realised by probabilistic methods, e.g. based on the repeatability of results and other parameters, and the systematic deviations.

## 5. QUANTITIES, SIGNALS AND THEIR PROPAGATION

GUM [7] allows both the propagation of uncertainties and the propagation of distributions. Classically, the expected value and the measurement uncertainty are calculated according to GUM [7] using the Gaussian uncertainty propagation. Thereby the measurement uncertainties of the input quantities and signals are represented probabilistically and are divided into type A and type B uncertainty contributions. Type A uncertainty contributions represent random deviations and are calculated by statistical evaluation of measurement series. Type B uncertainty contributions represent systematic influences due to the used equipment, environmental conditions, etc. [5],[11]. In Figure 6 [5] the concept of the GUM procedure [7] is illustrated.

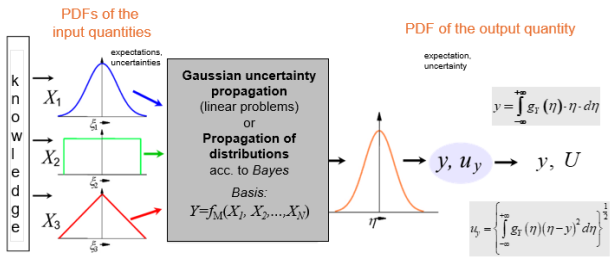

Figure 6. Illustration of the concept of the GUM [7] procedure [5].

For not normally distributed input quantities and nonlinear models the *Bayes-Laplace theory* can be used based on the propagation of PDF. In the Bayesian approach, a priori knowledge about the measurand is available, which is combined with the knowledge from the measurement to improve the knowledge about the measurand and refine the uncertainty assessment. Figure 7 shows simplified illustration of the Bayesian metrological estimation approach [12].

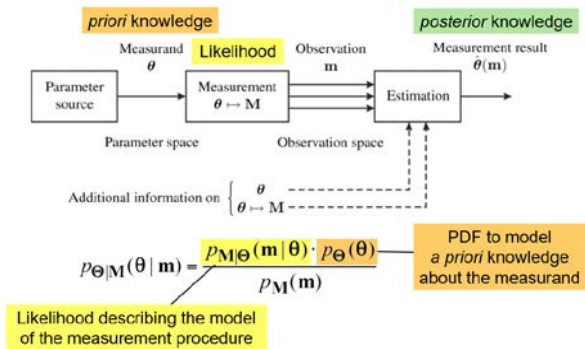

Figure 7. Simplified illustration of the Bayesian metrological estimation approach [12].

An appropriate way for not normally distributed input quantities and signals or if their influence on the output quantity is described by a non-linear function is the *Monte Carlo Method* (MCM). The basis of the MCM is the propagation of probability density functions of the input quantities by means of an experimental model. The probability distributions are obtained by applying the MCM coupled with appropriate definitions for the total measurement uncertainty. Figure 8 [13] illustrates the MCM as applied in metrology.

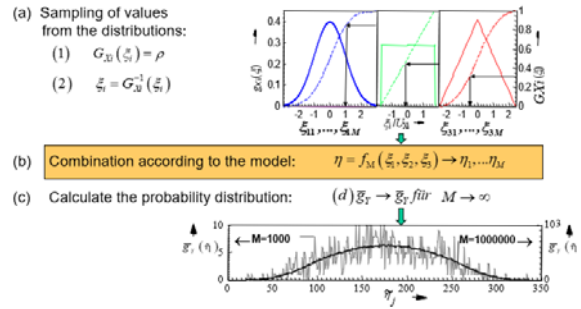

Figure 8. Bayesian approach based on MCM using integration techniques for propagation of distributions [13]

## 6. NETWORKED SYSTEMS: SENSOR DATA AND INFORMATION FUSION

Due to the growing importance and role of the digital transformation towards the *Internet of Things* (*IoT*) with an increasing number of intensively and flexibly networked measuring systems and consequently ever larger amounts of data to be processed, data-based modelling approaches have gained enormous importance in the last decade and will be used progressively more intensively. Important applications of networked systems include distributed process and production measurement, autonomous systems, climate observation, energy networks and security [14]. Figure 9 [10] illustrates a basic example of networked measurement and sensor system as they can be found in industry.

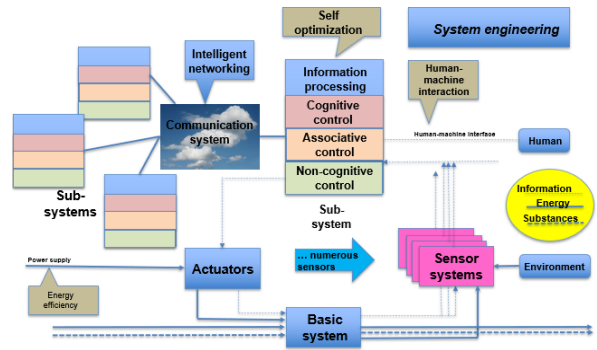

Figure 9. Example of a networked measurement and sensor system [10]

In networked systems consistently *grey box models* are used due to their ability to deal with these highly complex measuring systems for which the physical principles are not fully know explicitly. Additional challenges in networked systems arise from the modularity of the system itself. The combination of the sub-systems may depend on the quantity of interest, i.e., depends on the question to be answered from the networked system [15]. Hence, modelling approaches are often rather probabilistic than deterministic, i.e. *Bayesian*, *Dempster-Shafer* [16] and information-theoretical approaches. Sensors could as well be sub-systems of the sensor network.

Particularly relevant developments for measurement technology in networked systems are the *Digital Metrological Twins*, *Quantum Sensors* and *Self-X Approaches* (such as self-validation and self-calibration) based on redundant information, *Soft Sensor Technology* and, of course, high-performance *Data and Information Fusion*, available today [17]. The latter is particularly indispensable where similar or different sources of information must be consistently combined, i.e. merged to produce more accurate results.

### 6.1. BASICS OF SENSOR DATA FUSION

Information fusion is the merging, overlaying and utilisation of data and information contributions of several information sources with regard to a given task. The aim of information fusion is to favourably combine the contributions of the individual sources, which are not sufficiently informative on their own, in order to obtain more reliable or new knowledge [18]. For example, the fusion of observations from the sensors in a sensor network can be used to reduce measurement uncertainty, increase the detection range or minimise the risk of failure. Sensors provide data that contain information about the measured object. Information fusion combines this information of different sensors with the aim to get as much resulting information as possible [19][20].

There is particular potential in the fusion of heterogeneous information sources [20], as these often have different strengths and weaknesses and therefore complement each other. Consider, for example, the fusion of LiDar, ultrasound and radar measurements in autonomous driving. Examples include military or civilian reconnaissance, automatic visual inspection and cognitive robotics.

Bayesian probability theory, fuzzy theory and Dempster-Shafer's evidence theory are particularly suitable methodological approaches for data and information fusion. The main differences between Dempster-Shafer und Bayes are briefly outlined below [16]. A prerequisite for the application of Bayes' theorem is the existence of a set of complete and mutually exclusive individual hypotheses. In particular, alternative hypotheses must exist for each individual hypothesis. If this is assumed, the Dempster-Shafer theory and Bayes' theorem lead to the same result. In all other cases, Bayes' theorem is not defined. This gap is closed by the Dempster-Shafer theory as an extension of Bayes' theorem, in that vague information, for example the confirmation of several basic hypotheses by a single sensor, can also be meaningfully introduced.

Another disadvantage of Bayesian theory is the need for a priori knowledge about the hypotheses or classes with regard to their fulfilment. Such information can only be obtained very roughly from statistical analyses and is therefore often determined heuristically. The Dempster-Shafer theory does not require this explicit a priori knowledge as a matter of principle. It makes a clear distinction between the measurement results and its reliability. The disadvantages of the Dempster-Shafer theory are the comparatively high computational effort and the undefined nature of the combination rule in the case of completely contradictory statements [16].

### 6.2. DIGITAL TWINS

The intensive digital networking and data processing has already led to new approaches in measurement technology and industry. These include the so-called *Digital Twins* of both, measurement and process systems. According to ISO/IEC 30173 [21], a Digital Twin is a digital representation of a target entity with data connections to the real world. This digital representation contains selected characteristics, states and, depending on external influences, the behaviour of the measured object or process. It is therefore a system of aggregated digital models that are controlled by processed external data and information [22],[23]. Within this representation, different models, information, or data can be linked together during different life cycle or process phases [22].

As models of the represented measurement object or process and its behaviour, Digital Twins enable simulations with different, flexibly selectable target directions. They can contain algorithms and services that describe or influence the properties or behaviour of the represented object or process or offer services about them.

The concept of the Digital Twin and the Digital Shadow is based on computer models with associated inputs and uncertainties of the underlying object or process. Figure 10 shows the principal operation of a Digital Twin in measurement and sensor technology and metrology [2].

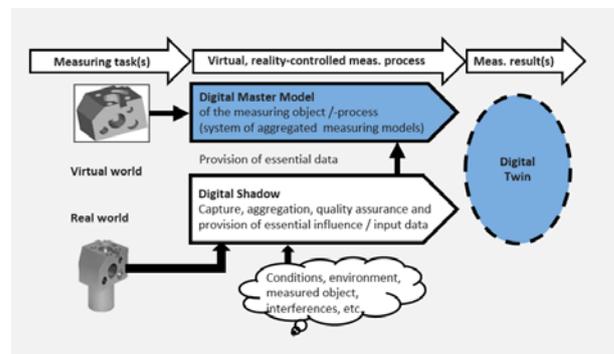

Figure 10. Principle function of a Digital Twin in measurement and sensor technology [2]

For a given measurement task, data from the real world and from the digital representation are combined (fused) to form the Digital Twin of the measurement. The Digital Shadow may already contain fused data and information, partially from networked measuring systems. The action is then again combined with measurements from the real world to update the Digital Twin.

### 6.3. QUANTUM SENSORS

The revised SI means that the units, which are now based on seven so-called *Defining Constants*, can be represented in different ways using (physical) realisations. Now one-step traceability to these Defining Constants of the *Revised International System of Units* is possible. It reduces the effort required for recalibration by "breaking" the classical "Calibration Chain". The redefined International System of Units now enables recourse to reliable values based on natural constants. This change represents a major advance in metrology. It means that the Defining Constants are exactly i.e. uncertainty-free known by their definition. It is therefore now possible to fall back directly on reliable values. Only functional uncertainty of the sensors is to be taken into account which e.g. might be evaluated by comparison measurements. For example, the National Institute of Standards and Technology (NIST) has launched *NIST on a Chip*, a program that aims to make ultra-reliable measurement technology available nearly anywhere and anytime.

## 7. COGNITIVE SENSORS AND MEASURING SYSTEMS

Cognitive systems, also known as AI-driven systems, are logical developments from Machine Learning (ML), the theory and application of neural networks. From the metrology point of view, cognitive / data-driven systems might be categorised as grey box models, as sound prior knowledge must always be available (learning process). An exemplary topology of a neural network is shown in Figure 11.

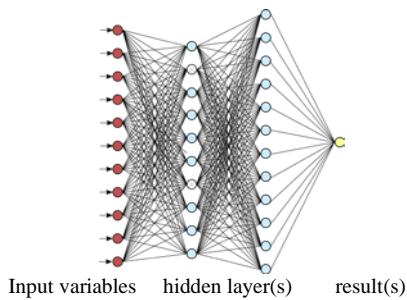

Figure 11. Exemplary topology of a neural network

First "bloom" of cognitive sensors and measuring systems in measurement technology was in pattern and image recognition, recognition of disease patterns in medicine, in some cases superior to traditional medicine, speech recognition, etc. The main advantages that data-based AI-driven sensing and measurements can be expected to offer or are already offering are:
- possibility of virtual sensing and realization of so-called soft sensors
- realization of cost-effective sensors using data-based strategies to estimate results of complex measurements
- prediction of sensor or measurement system behaviour, e.g. degradation or temperature dependency (cf. section 5.2)
- reduced measurement uncertainty using data-driven approaches
- handling of complex high-dimensional data, e.g. for sensor data fusion (cf. 5.1)
- more efficient measurement data processing using machine learning and data processing pipelines etc.

### 7.1. TRUSTWORTHINESS AND EXPLAINABILITY

Despite the existing and tempting advantages, there are still significant risks and uncertainties of these approaches existing. With regard to AI-driven systems, all these imponderables are summarized under the terms trustworthiness or - negatively – risk which cannot be clearly interpreted metrologically yet.

At present, explainability is the core of trustworthy ML and uncertainty quantification, as it supports building trust in model predictions and aids in identifying and managing uncertainty sources. Therefore, for end-users having access to an uncertainty evaluation and explainability solution, this is essential for building trust and increasing acceptance of the algorithm.

In addition, explainability strategies can also lead to a reduction of ML decision layers and thus of (metrological) sources of uncertainty. Explainability therefore forms a bridge to the so-called white box models, i.e. to our knowledge of the strategy, physics, chemistry, etc. of our measurement or measurement system.

### 7.2. UNCERTAINTY QUANTIFICATION IN MACHINE LEARNING

Uncertainty quantification in ML and AI-driven systems identifies and quantifies uncertainty in data, models and algorithms, serving as a tool to assess AI trustworthiness. Since AI predictions are impacted by uncertainties, reliable estimates are crucial for providing diagnostic insights for both, developers and users and for understanding the system limitations. Such estimates guide improvements in data representation, data collection, and model tuning, and can reveal potential biases, deviations, or inconsistencies in decision-making processes. In high-risk applications, where high-confidence explanations are essential, quantifying uncertainties in explanations is crucial [24],[25].

Sources of uncertainty in ML and AI usually are divided into two categories [25]-[30]:
- *Aleatoric (statistical) uncertainty* refers to the notion of randomness, that is the variability into the outcome of an experiment due to random effects.
- *Epistemic (systematical) uncertainty* refers to uncertainty caused by a lack of knowledge, i.e., to the epistemic state of the quantity.

Aleatoric uncertainty accounts for the natural variation of inputs and parameters and refers to inherent variability or noise in the data itself. Therefore, we may also say that aleatoric uncertainty, also called data uncertainty, refers to the statistical uncertainty. It is related to randomness where a larger amount of data cannot reduce the randomness of the outcome [26],[27]. Aleatoric uncertainty can arise from various sources, such as measurement errors like random noise, but also due to insufficient information, e.g. low resolution or imperfect measurements [28].

Epistemic or model uncertainty refers to the uncertainty due to a lack of knowledge or understanding about the ML model or the data. Epistemic uncertainty can arise from various sources, such as limited or biased data, data shift/drift - change in real-world data compared to the training data, insufficient training data, model architecture decision - uncertainty can also arise from the design choices, also called inductive bias, of the deep learning model architecture/setup. This type of uncertainty is reducible by more evidence, which means in the field of machine and deep learning by more information / data. Epistemic uncertainty can stem from multiple sources, either from a data or modelling perspective [27]. Epistemic uncertainty from the data perspective can arise due to dataset shifts, also known as distribution shifts. Distribution shifts refer to the change in real-world data compared to the training data, violating the default assumption in machine learning that the train and test data are independently and identically distributed [27]. This can stem from changes in the environment overtime, such as changing weather conditions but also from the high variability and inherent complexity of real-world environments. The training data may not adequately cover this total variability, which can result into an imbalance in the training data from an unknown sampling selection bias where not all populations are equally represented [27]. Aleatoric versus epistemic uncertainty in ML is shown Figure 12 [26].

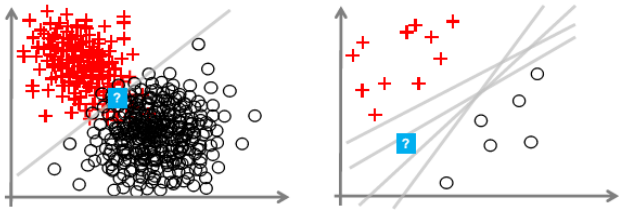

Figure 12. Left: Even with precise knowledge about the optimal hypothesis, the prediction at the query point (indicated by a question mark) is aleatorically uncertain, because the two classes are overlapping in that region. Right: A case of epistemic uncertainty due to a lack of knowledge about the right hypothesis, which is in turn caused by a lack of data [26].

Thus, there is a major correspondence between aleatoric uncertainty and type A uncertainty in metrology as well as epistemic uncertainty and type B uncertainty, but no complete correspondence. Above all, the statistical or probabilistic incorporation and propagation of uncertainties is not defined in the field of ML and AI.

Uncertainty occurs in various facets in machine learning, and different settings and learning problems will usually require a different handling from an uncertainty modelling point of view [29].

In Figure 13 the relation on model induction, predictive model and new query instances in ML is illustrated [29]. This is comparable with the Bayesian approach in Figure 5.

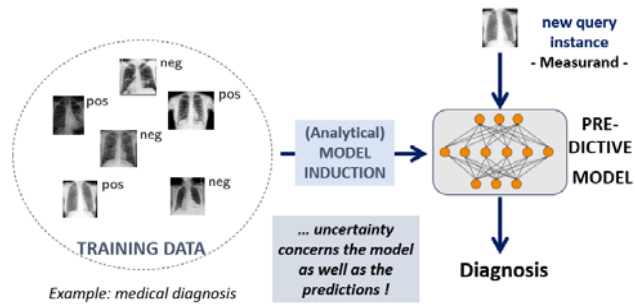

Figure 13. Uncertainty combination in predictive / cognitive measuring systems. [29].

The basic setting of supervised learning a hypothesis is induced from the training data and used to produce predictions for new query instances [29].

### 7.3. AI-DRIVEN SYSTEMS FOR MULTIPLE SENSOR DATA FUSION

Artificially intelligent automation has not only impact on sensor technologies, but also on comprehensive multiple sensor systems for assisting situational awareness and decision-making [18]. From a systems engineering perspective three tasks need to be fulfilled [18]:

(1) Design artificially intelligent automation in a way that human beings are mentally and emotionally able to master each situation.

(2) Identify technical design principles to facilitate the responsible use of AI in ethically critical applications.

(3) Guarantee that human decision makers always have full superiority of information, decision-making, and options of action.

This requires addressing the following issues [18]:
− algorithms needed,
− the data to be processed,
− the programming skills required,
− the computing devices to be used,
− the inevitable anthropocentric design,
− the reviewing of R & D effort necessary, and
− the integration of different dimensions in a systems-of-systems point of view.

### 7.4. IMMEDIATE METROLOGICAL CHALLENGES

The creation of AI-driven measurement systems is a reality. A number of scientific efforts are currently still required to integrate these into the metrological hierarchy: These are in particular:
− clarifying the semantics of uncertainty concepts in ML and AI,
− clarifying the white, grey and black box semantics (gravity point in AI, ML: interpretability of the model) and
− their clear definitions encompassing both areas, metrology and ML.

In particular, the learning processes of ML systems must be examined metrologically. Completely new fields for metrology, however, are the explainability of systems and the treatment of important human intervention, e.g. in medical applications.

### 8. CONCLUSIONS

In this paper we have discussed classical methods of modelling in metrology like analytical parametric, data-based modelling and grey box modelling.

The present, and certainly the future of data-based modelling, will be increasingly dominated by intelligent, cognitive systems. Here, we are only at the very beginning of evaluating the accuracy, trustworthiness or even the functional uncertainty of the models.

We have taken the first steps towards explainability and risk assessment, but it seems clear today that in the end there will be a probabilistic, uncertainty-equivalent quantitative evaluation, possibly, e.g. based on the repeatability of the training results and the systematic deviations. Functional uncertainty can be seen as a common anchor between the different modelling approaches.

The advantages of the presented data-based and grey-box modelling are that no explicit physical understanding of the model is required and modelling and fine-tuning of the model during operation is possible.

New approaches that are important for modelling and should be further developed are *Digital Twins, Quantum Sensors* and *the Fusion of Multiple Sensor Systems* for assisting human situational awareness and decision making.